# UNDERSTANDING THE EMERGENCE AND DEPLOYMENT OF "NANO" S&T

Barry Bozeman

Ander Crenshaw Chair and Regents' Professor Public Policy
Dept. of Public Administration and Policy
201 Baldwin Hall
University of Georgia
Athens, Georgia 30602-1615
USA
Mail: bbozeman@uga.edu

Philippe Laredo

Université de Paris-Est (ENPC)
and Manchester Business School
Cité Descartes, 77455 Marne la Vallée cedex 2, France
Mail : philippe.laredo@enpc.fr

Vincent Mangematin

GAEL (INRA/UPMF) and
Grenoble Ecole of Management
12 rue P. Sémard
38003 Grenoble Cx
France
Mail : Vincent@mangematin.org

Abstract

As an introduction to the special issue on "emerging nanotechnologies", this paper puts in perspective contemporary debates and challenges about nanotechnology. It presents an overview of diverse analyses and expectations about this presumably revolutionary set of technological, scientific and industrial developments. Three main lines of argument can then be delineated: first of all, the degree of cumulativeness of science and technologies and the respective roles of newcomers and incumbents in the industrial dynamics; second the knowledge dynamics in nanotechnologies, especially the linkages by science and technology and third the role of institutions (network, geographic agglomeration and job market). It finally discusses methodologies to delineate the field of nanotechnologies and to collect data.

The focus of this special issue is on the creation of new science-based industries. It considers the pursuit of scientific and technical breakthroughs pertaining to matter at the level of the nanometer, as explained in the definition proposed by the US National Nanotechnology Initiative, a definition fast becoming the standard. (See box).

> National Nanotechnology Initiative's (NNI) definition:[1]
> "Nanotechnology is the understanding and control of matter at dimensions of roughly 1 to 100 nanometers, where unique phenomena enable novel applications. The diameter of DNA, our genetic material, is in the 2.5 nanometer range, while red blood cells are approximately 2.5 micrometers. Encompassing nanoscale science, engineering and technology, nanotechnology involves imaging, measuring, modelling, and manipulating matter at this length scale.
> At the nanoscale, the physical, chemical, and biological properties of materials differ in fundamental and valuable ways from the properties of individual atoms and molecules or bulk matter. Nanotechnology R&D is directed toward understanding and creating improved materials, devices, and systems that exploit these new properties."

Nanoscience and nanotechnology research (hereafter "nanotechnology") appear to have the potential to revolutionize many sectors of industry, in particular by fostering the convergence between previously distinct technology-driven sectors. The proponents of such massive transformation have coined a set of interconnected terms to qualify the nature of the anticipated changes: they speak of the convergence of NBIC meaning Nanotechnology, Biotechnology, Information and communication technology, and Cognitive sciences. Some use the evocative acronym "BANG," implying the interconnectiveness of, respectively, Bit, Atom, Neurone and Gene. Nanotechnology expectations are not solely about industrial innovation but the creation of a generic industry that will penetrate and transform other industries. Exhibiting the epitome of creative destruction, nanotechnology advocates claim that nanotechnology will redefine existing industries and array them in new combinations, changes being already underway, as submicronic technologies entangle communication and information industries.

Contributions in this special issue contribute to a venerable tradition of technology and industrial change theories. They have been inspired by the work of Thomas Kuhn on the structure of scientific revolutions (Kuhn, 1962) or scholars such as Dosi (1982), Anderson, and Tushman ( 1990) and Afuah and Utterback (1997) who suggest that both the cognitive conditions and the industrial structure by which knowledge is generated change in response to the maturity of underlying technologies. Life cycle theorists identified two major phases which present a deep internal coherence in the way knowledge is being produced (Abernathy and Utterback, 1978; Tushman and Anderson, 1986; Utterback and Suarez, 1993). The first phase is characterised by rapid technical change whereas the second phase organises technological consolidation around a dominant design (Anderson and Tushman, 1990). The emergence phase opens with the introduction of a radically new invention, which is believed to be superior to the existing one in terms of production costs, productive services and market opportunities. Far from providing a stable and economically superior set of solutions, the exploitation of the new technology calls for further exploration: scientific and technological hypotheses have to be tested one against another. Hence the exploration of competing technological hypotheses or trajectories takes place in a turbulent environment, where the introduction of new technical solutions amplifies uncertainty instead of reducing it.

---

[1] NNI is a U.S. federal R&D program established to coordinate the multi-agency efforts in nanoscale science, engineering, and technology. .See http://www.nano.gov/html/facts/whatIsNano.html

The arrival of a radically new technology generates new firms, either in existing industries or entirely new ones. These new firms are based on their distinctive technical skills and can be viewed as independent research projects exploring competing technological opportunities. Hite and Hesterly (2001) point out that during the early stages of a new industry, start-ups face great uncertainty about the efficiency of their ill-defined routines and products and how these fit the environment within which they are embedded. Scientific activities remain close to the edge of knowledge; the knowledge produced incorporates a large proportion of tacitness and remains embodied in those who produce it. It follows that the circulation of knowledge equates with the circulation of researchers or engineers themselves (Almeida and Kogut, 1999; Bozeman and Mangematin, 2004). This does not deny the fact that in the meantime, incumbent firms may equally invest in the new technology. However as emphasised repeatedly, technology acquisition is far from immediate and gratuitous. Instead, it requires long-term investments in knowledge acquisition. In recent time, especially after the 1980's revolutionary developments in computers and information technologies, revolutionary technology opportunities motivated large firms to establish ties with diverse research institutions, especially government laboratories and universities, and to participate in research consortia.

This introduction tries to set the intellectual scene and put in perspective contemporary debates and challenges about nanotechnology. While all the papers in this issue deal with the emergence of nanotechnologies, the authors provide diverse views and expectations about this presumably revolutionary set of technological, scientific and industrial developments. Three main lines of argument can then be delineated: first of all, the degree of cumulativeness of science and technologies and the respective roles of newcomers and incumbents; second the knowledge dynamics in nanotechnologies and finally the role of institutions.

## The Respective Roles of Newcomers and Incumbents

Zucker et al. (2007) propose that science and technology broadly defined are cumulative. The data gathered show that the capacity of producing in a given space and in a given domain a new piece of knowledge is strongly correlated to the pre-existing stock of codified knowledge (here articles and patents) in this space and in all domains, even when faced with a rapid rate of knowledge obsolescence. They thus emphasise geographic proximity, high tech firm creation and circulation of tacit knowledge through human resources, similar to the recent history of biotechnology.

Hill and Rothaermel (2003) underline that tacit knowledge and know how are embedded in routines of existing organisations. Thus routines of incumbents cannot change rapidly. On the contrary, new comers introduce new routines in the industry. Numerous works have highlighted the central role science based start-ups in biotechnology, including research studies by authors in this special issue (Corolleur et al., 2004; Rothaermel and Thursby, 2005; Zucker et al., 1998). And there has already been a number of publications studying start-up firms in nanotechnology (Porter et al., 2006). However, when considering biotechnology, as mentioned by Rothaermel and Thursby (2007) previous assumptions are being reconsidered due to the longer time frame and the longitudinal studies it enables. In particular it is now assumed that even in biotech, there has been relatively little displacement of incumbents, pushing forward the idea that "creative destruction" took place within existing large firms (especially pharmaceutical firms). This does not discard the role of start-up firms, but requires a repositioning with respect to our understanding of the dynamics of an emerging field. Some authors (Mustar et al., 2006) argue that they fill a "knowledge gap" by embedding the new knowledge produced into instruments (such as the AFM and STM microscopes, and more and more new modelling and design tools in nanotechnologies) and by demonstrating to users and stakeholders the wider value of the breakthrough technology through positioning in niche markets. This remains to be studied in more depth and is an important line of research we advocate for the future.

Two papers in this special issue focus on incumbents. Rothaermel and colleagues (2007) examine key factors explaining the performance of firms over time, while Avenel et al. (2007) delineate knowledge building strategies of firms. The latter researchers confirm a convergence of the nano knowledge base of firms through the examination of the technological diversity of both patents and the patent portfolio of firms. However, they show that small and large firms deploy different strategies: convergence is at the level of individual patents for small firms, while it is mostly at the level of the patent portfolio for large firms. This might well confirm the previous hypothesis on the role of small firms in the overall dynamics of this emerging industry. However it leaves open the reasons why large firms invest so heavily, far beyond what may be considered necessary to fuel firms' absorptive capacity. Avenel and colleagues' research suggests two trajectories of innovation, through hybridizing the existing knowledge base for large firm and via the exploitation of breakthrough knowledge in small firms.

These findings link to a key result from Rothaermel and Thursby (2007). In their article, the authors test a model where knowledge performance (as measured by the number of granted patents) depends on the articulation between the internal knowledge base of firms (measured by their lagged patent portfolio) and their absorptive capacity (the markers of which are R&D alliances and/or acquisitions of R&D intensive small firms). Looking at biotechnology and pharmaceutical incumbents over 20 years, they demonstrate the initial critical role of alliances while a shift occurs in favour of internal R&D investments, when techniques and instruments become commercially available. With these results in mind they turn to a wider set of incumbent firms which have at least patented once in nanotechnology over the same period of time. They show widely different results where internal R&D investment is the critical factor, associated in the recent period (after the commercialisation of key instruments on a wide scale), with a significant role of acquisitions. They suggest that nanotechnology is at a different degree of maturity in the technology life cycle. They argue that initial enabling technology only fostered research activities and that we are still in need of "new methods for inventing" (Darby et al., 2003) for fostering innovations. Robinson et al. (2007) suggest that these might also depend on further deployment of "research facilities and technological platforms." Until these platforms fully deploy, we remain at a pre-alliance stage focusing on the accumulation and testing of new knowledge by actors.

## Knowledge Dynamics

The industrial dynamics of emerging nanotechnologies as well as the formation of alliances depend upon how the search process is organised and whether it is of the same nature as previous waves, including IT and biotech. Bonaccorsi and Thoma (2007) building on recent work on search regimes (Bonaccorsi, 2005), propose a revised "industrial economics of research." Their model is based on three properties which characterise knowledge production: the rate of growth, the number of options and directions which the researchers explore (and thus the degree of cumulativeness of knowledge), and the nature of complementarities required. These requisite complementarities may be cognitive (e.g. interdisciplinarity), technical (e.g. big science and large facilities such as ITER) or institutional (mixing producers from different institutional background, e.g. university-industry). Using the same sources as previous papers (articles and patents), they argue that nanotechnology witnesses over the last 15 years a very rapid rate of increase, far above the growth of science and technology at large (with striking figures for articles: 14% against an average of 2%) and that it is a highly diverging science (using as a marker the yearly rate of appearance of new keywords – consistently over 40% for the whole period).

They then focus on institutional complementarities, based on the more than 8000 inventors retrieved in their database. They propose a simple taxonomy which drives to striking results. Separating inventors on the basis of publications in academic journals, they classify patents in three classes: patents which gather inventors that are "inventors only", patents in which all inventors are also "authors" of academic articles, and patents that mix the two. This drives them to confirm the suggestion of Rothaermel and Thursby that nanotechnology is best viewed as being at a very early stage, inasmuch as two-thirds of patents have at least one inventor who has also published academic articles. Looking then at performance (the highest patenting authors) and at quality of patents (mixing diversity, breadth and extension), they show that hybrid "author-inventor" patents are the most promising, underscoring the critical importance of institutional complementarities.

This latter result is in line with other studies (Dietz and Bozeman, 2005) and with the second point by Zucker et al. (Zucker et al., 2007) about the "geographical localisation of knowledge" and the key role of "cross-institutional linkages." While Bonaccorsi and Thoma (2007) demonstrate the importance of heterogeneous linkages in nano-patenting, Zucker et al. (2007), when looking at the relative performance of the 179 US functional economic areas, show the importance in nano production of previous cross-institutional links (seen through the lenses of co-authorship and co-invention). Both results could thus be complementary: the more pre-existing cross-institutional channels at the "area" level, the more chances to develop "author-inventor" patents, and the more productivity of localised actors.

When enlarging the analysis at the country level, these results pertain to the warning made by Guan (2007) when analysing the rapid growth of Chinese publications in nanotechnology. China is the now the second country for the number of publications, but over 99% of such publications are produced by universities and the Chinese Academy of Science.

The explanation underlying the role of cross-institutional linkages and localised knowledge production, deals with the importance of tacit dimensions in emerging, fast growing and yet fluid areas of knowledge. The stickiness of knowledge would produce "temporary natural excludability" (Rothaermel and Thursby, 2007) driving to multiply cooperations and/or alliances in order to overcome them. At the same time the uncertainty that prevails in breakthrough science requires numerous face-to-face debates and mutual adjustments. Thus, there is an advantage to geographical proximity, above other forms of proximity (Boschma, 2005). Knowledge agglomeration is once again manifest.[2]

However, considering the high level of clustering in biotechnologies, where 50% of the US biotech firms were created within four geographic areas (Powell et al., 2002), one wonders whether co-location is specific to nanotechnology. Does nanotechnology follow evolutionary paths similar to earlier technological waves? Or does the nanotechnology wave exhibit unique patterns? Robinson et al. (2007) suggest that a major difference deals with how to manipulate and produce at the nano-scale and with the key role of research facilities. Their assertion is that "technological agglomeration is the effect of technological platforms being set-up, used and expanded." This then strongly articulates knowledge dynamics with the third line of argumentation, focusing on institutional transformation.

### The role of institutions

The above-mentioned articles have already demonstrated that institutions matter. By insisting on "cross- institutional linkages" they resemble previous works on innovations systems, and more specifically on regional systems and clusters. However they only focus on linkages, channels or networks not addressing the organisations themselves. Jong (Jong, 2006) demonstrates in his study of the San Francisco Bay area that the "research environment" at one university, University of California- San Francisco, played a central role in the creation and growth of biotechnology firms. This research environment is at the core of the comparative analysis, proposed by Robinson et al. (2007), of nanotechnology research and industry in Grenoble, France and Twente, The Netherlands. They demonstrate the importance of organisational transformations entailed by the respective creation of Mesa + and Minatec, and question the factors underlying the two very different routes taken. Whatever the differences, a critical dimension is the changing conditions under which research activities are undertaken, based upon the growing role of technology platforms and "long distance" interdisciplinarity.

---

[2] This argument remains to articulate with the stream of literature on localised knowledge spillovers following

This links with the result obtained by Stephan et al. (2007) in their study of nanotechnology training at universities. They show that "on the job" training via research activities dominates, highlighting the critical importance in training, of "ways of doing" nanotechnology research (Pickstone, 2001). However their survey of academics in nanocenters drives them to a different interpretation: the absence of rotation in doctoral training (an important factor in biotechnology learning) emphasizes a "Principal Investigator"-dominated approach, typified by the development of research groups centred on the faculty members who have generated funding via grants and contracts. These findings may be at odds with those of Robinson et al. (2007).

The question remains- what is the role of the research and production facilities in nanoscale work? Are they a key ingredient fostering "technological agglomeration" and "convergence"?

## Further data requirements

When a field is fluid and crossing numerous borders (physics, chemistry, biosciences, and engineering), analysts face special challenges in capturing the field's dynamics. This is visible in this issue where five of the seven papers propose different delineations of the codified nanotechnology knowledge base (papers and patents). Other delineations have been produced recently (Noyons et al., 2003) or Kostoff (Kostoff et al., 2006a; Kostoff et al., 2006b). Similar efforts are being presently undertaken by the National Science Foundation center at Arizona State University (with a team headed by D. Guston) and companion work at Georgia Institute of Technology Tech (with a team headed by P. Shapira) and work headed by Zucker ("nanobank" project). On the methodological side, Zitt (2005) has investigated citation-based

---

Jaffe's seminal work.

methods to test the relevance of approaches based on keywords. This citation-based approach faces two strong limitations because of the time lag associated with the use of citations (an important issue when the field is both fluid and fast growing) and because of institutional aspects (linked to conditions of use of the WoS). This is why Mogoutov and Kahane (2007) propose a complementary approach fully based on keyword analysis which enables periodic updating, including the appearance of new keywords and associated with an open space to organise enrichment.

## This issue as a first step: Numerous alleys to further explore

Authors in this issue raise as many questions as they resolve, including the role of incumbents and of start-ups, the types of cross-institutional linkages which matter and the need for organisational change. The emergence of nanotechnologies questions the established technology life cycle theories and the impact of breakthrough innovations on industrial organisation. It opens avenues for further research on firm strategies in presence of technological breakthroughs. What are the relevant innovation strategies? What are the respective roles of incumbents and start-ups in the innovation process? How is industrial organisation transformed? How do these innovation processes shape new and old markets? Will we face the usual bilateral relationship between producers and users? Will the 'collective' intervention focus on classical issues of standardisation (Blind, 2006)? Or will we witness stronger regulation, pushed by controversies and citizens debates on the need for controlling and framing the deployment of nanotechnology products? Will this drive firms to focus on "embedded" nano solutions (as in chips)? Will it push public authorities to intervene and, for instance, apply a "pharmaceutical-like" regulatory approach where each new product requires a legal approval before being commercialised? These questions are far from being 'out of the box' conjectures when one witnesses initiatives by insurance firms (like Swiss Re Porro, 2004) or the growing controversies on all the ethical, public safety and participatory aspects of nanotechnology. The fact that democratic debate takes place before the event and during the exploratory stage is not completely new in the relationship between science and society (witness early participation in genetics engineering policy), but there is only minimal previous experience in handling the social complexities accompanying such revolutionary science and technology. The implications of nanotechnology for social change is already a strong focus for research as illustrated by the NSF centre for nanotechnology in society (Arizona State University) or the Dutch social science component of the Nanoned programme.